# 50 years INR: Exploring the High-Energy Universe

Christian Spiering

*Deutsches Elektronensynchrotron DESY*
*Zeuthen, D-15738, Germany*

**June 19, 2021**

**Abstract:** This article is an attempt to review 50 years of high-energy cosmic particle physics at the Institute for Nuclear Research of the Russian Academy of Sciences. It is written by an outsider whose scientific career, to a large part, was formed by collaborating with INR scientists in the late 1980s and 1990s. The review covers the fields of cosmic-ray, gamma-ray and high-energy neutrino physics. The main focus will be on INR's large infrastructures in the Baksan Valley and at Lake Baikal. Research at these infrastructures is flanked by participation in top experiments at different places around the world, recently the Telescope Array in the USA and the LHAASO detector in China.

## 1. Introduction

Cosmic radiation has been discovered more than hundred years ago. With a high-altitude flight in 1912, the Austrian physicist Victor Hess proved that some kind of ionizing radiation enters the Earth's atmosphere from outside. It took until the 1920s before it became undisputed that cosmic radiation consists of charged particles and not gamma rays. First visualization of single cosmic particles was achieved in 1927 by Dmitri Skobeltsyn in Leningrad, using a cloud chamber. The 1930s saw the discovery of positrons, muons and charged pions in cosmic rays. In 1939 the French physicist Pierre Auger discovered that cosmic rays at the surface of the Earth arrive in the form of particle showers. From the horizontal extension of the showers he estimated the maximum energy of the showers to be of the order of $10^{15}$ eV. Meanwhile we know that the maximum energies are actually much higher, in excess of $10^{20}$ eV. This raises the question of the origin of cosmic rays: how manages nature to accelerate particles to such mind-boggling energies?

Principally, the question can be addressed in three ways:

1) Measuring direction, energy and mass composition of charged cosmic rays.
2) Measuring direction and energy of high-energy gamma rays which have been generated in the source region by processes like $p + target \rightarrow \pi^0 + \cdots$ and the subsequent decay $\pi^0 \rightarrow \gamma + \gamma$ .
3) Measuring direction and energy of high-energy neutrinos which have been generated in the source region by processes like $p + target \rightarrow \pi^+ + \cdots$ and the subsequent decays $\pi^+ \rightarrow \mu^+ + \nu_\mu$ and $\mu^+ \rightarrow e^+ + \nu_e + \bar{\nu}_\mu$ .

Each method has its advantages and drawbacks. Cosmic rays as primary particles had raised the question. One can measure their energy and – to a certain degree – their mass composition. However, except of the very highest energies, their deflection in galactic magnetic fields is so strong, that directional information is lost and pointing becomes impossible. Gamma rays, on the other hand, propagate straight. But while 99% of cosmic rays are hadrons, gamma rays do not provide a water-tight proof of a hadronic origin. This is because they could also have been created by inverse Compton scattering $e^- + \gamma_{\text{low energy}} \rightarrow e^- + \gamma_{\text{high energy}}$, i.e. following a process of electron (not hadron) acceleration. Both charged cosmic rays and gamma rays of highest energies have a limited range since they can be absorbed by interactions with the cosmic microwave background. Neutrinos, at the end, do not suffer from these drawbacks, they travel straight and are not absorbed. However, their small interaction cross section makes them extremely difficult to detect.

In the following, I will describe the history of cosmic-ray, gamma-ray and neutrino astrophysics at INR. Due to my own specialization in neutrino astronomy and my former membership in the Baikal Collaboration, there will be some disbalance between sections 2 and 3 on the one hand and section 4 (neutrino astronomy) on the other. I beg the pardon of the reader for that somewhat subjective and selective character of the review.

Before starting the walk through 50 years of history, let me reproduce the portraits of the three physicists which can be considered fathers of cosmic-ray physics at INR and whose names will appear several times thorough this text: Mosei Markov, Georgi Zatsepin and Aleksandr Chudakov.

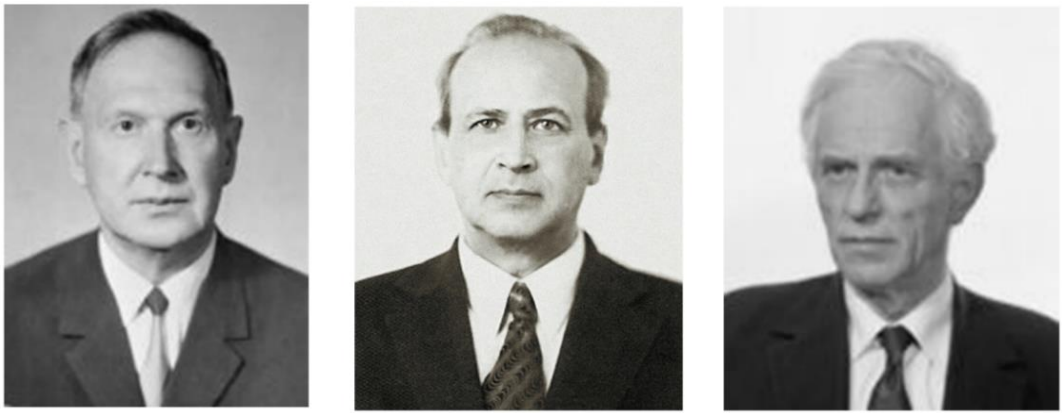

**Figure 1:** *Fathers of cosmic-ray and neutrino physics at INR, from left: Mosei Aleksandrovich Markov (1908-1994), Georgi Timofeyevich Zatsepin (1917-2010) and Aleksandr Evgenievich Chudakov (1921-2001).*

## 2. Cosmic Ray Research

### 2.1 The Baksan Neutrino Observatory (BNO)

The Baksan Laboratory of INR is mostly known for its underground part with the Baksan Underground Scintillation Detector (BUST) and the SAGE experiment as the main detectors. These two detectors focus on low energy and atmospheric neutrinos [1]. Yet, BUST plays also a role for cosmic ray physics. The detector consists of four horizontal and four vertical layers of liquid scintillation counters [1]. It is located 300 m under the surface and was commissioned in 1977.

Nineteen years later, in 1996, the Andyrchy air-shower array, located on the surface above BUST, started operation. It consists of 37 scintillation counters [2]. Figure 2 shows the configuration of the two detectors. The combination of a surface array and an underground detector allows measuring the air shower particles (mostly electrons and positrons) on the surface and the muons from the decays of secondary particles underground.

---

[1] See section 4 on atmospheric neutrinos in BUST, and the Addendum on solar and supernova neutrinos.

The first large facility of the Baksan Laboratory, however, was Carpet (ковер), located about 600 m from to the entrance of the Underground Lab. This air shower array was constructed in 1973 and consists of liquid scintillator stations. Its central part (the "carpet") consists of 400 scintillators with a total area of 200 m² and served as a model of a single layer of the future BUST. Stations at the surface record the full electromagnetic content of the shower, stations at shallow depths (which later have been added) serve as muon counters [3].

The initiator of Carpet and BUST was Aleksandr Chudakov, while upgrades, operation and analysis in the 1980s and 1960s were coordinated mainly by Alexander Voevodsky and Evgenij Alexeyev.

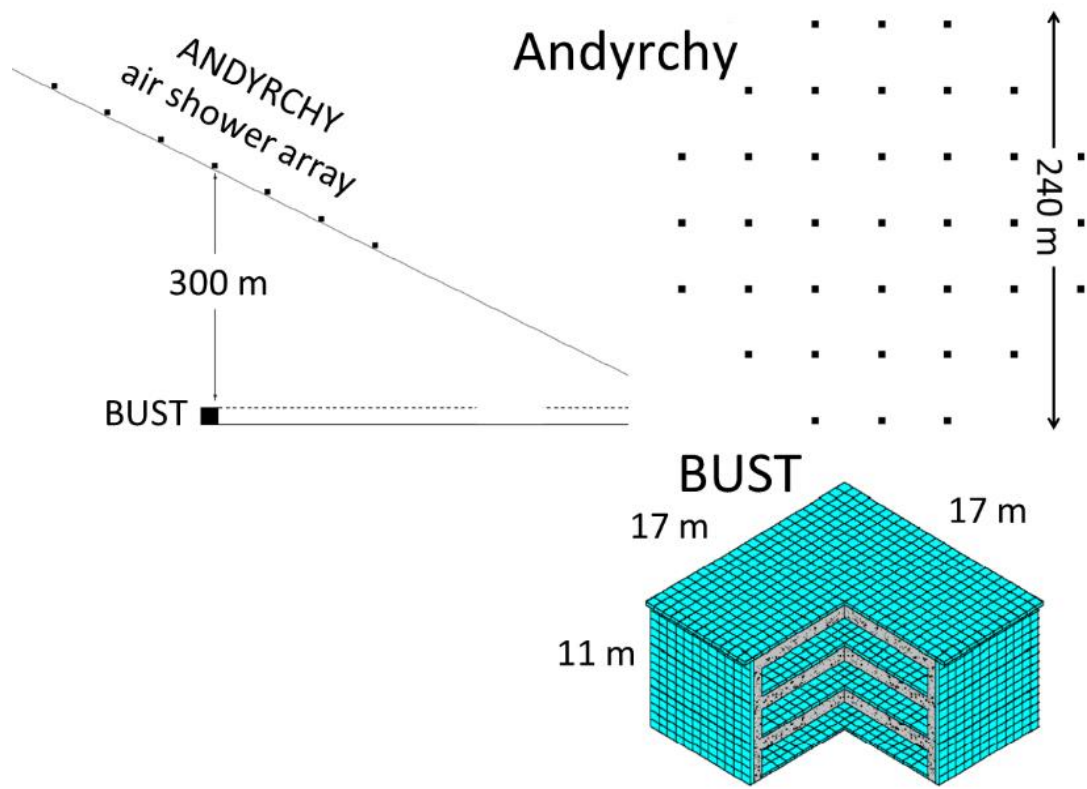

*Figure 2: The Baksan air shower array ANDYRCHY and the Baksan Underground Scintillation Telescope (BUST).*

From mid of the 1970s on, Carpet delivered high-precision lateral distribution functions of extensive air showers, measuring also showers with multiple cores from jets with high transversal momenta. That allowed Carpet deriving the cross section for the generation of high-$p_T$ jets at a C.M.S. energy of about 500 GeV. In [4] it was demonstrated that the cross-section of high-$p_T$ jet production, derived from the analysis of multi-core showers in Carpet, was in a good agreement with QCD predictions. One and a half year later the CERN SPS-collider started operation and the Carpet result was confirmed (of course with much better accuracy) by the UA1 and UA2 experiments (see Fig.3 and also [5]).

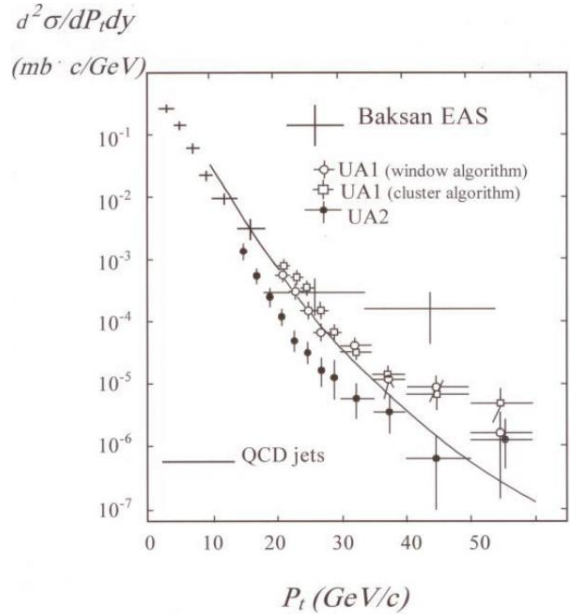

*Figure 3: Cross section of high-$p_T$ events in Carpet (1981) compared to QCD predictions and to first results from UA1 and UA2 experiments (1982).*

Also BUST (mostly known for its co-discovery of neutrinos from the supernova SN1987 A) has delivered a large number of important results in cosmic ray and particle physics. This includes the indirect measurement of the energy dependence of the photo-nuclear ($\gamma$N) interaction cross section, derived from the study of inelastic muon interactions [6]. The energy spectrum of high-energy muons generated in air showers [7] or the measurement of the multiplicity of muons and its relation to the mass composition of primary cosmic rays [8]. Figure 4 (taken from [9]) shows the photon-proton cross sections derived from the BUST data [6] together with approximations of photon-proton cross sections derived via the vector-meson dominance model from pion-proton data (upper curve) and nucleon-proton data (lower curve). The data in between are direct photon-proton cross sections at lower energy and derived from electron-proton collisions (ZEUS, H1) at higher energy (lines are fits with two different models). For the study of the primary spectrum of cosmic rays and the composition around the knee, the combined operation of the Andyrchy array and BUST was of particular advantage (see e.g. [10]).

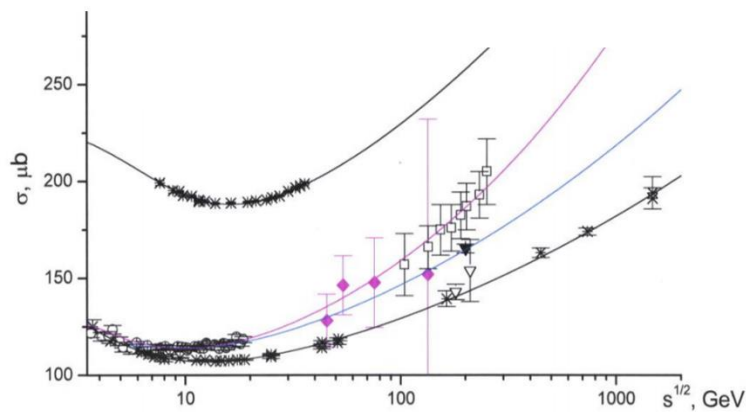

*Figure 4: Energy dependence of $\gamma$p interactions [9]: The BUST results are marked by the ◆ sign. See text for further explanations.*

## 2.2 The Telescope Array (TA)

The Baksan detectors address the energy range between $10^{13}$ and $10^{16}$ eV. To study higher energies, much larger air shower arrays are necessary. The presently leading experiments for the highest energies of cosmic rays are the Pierre Auger Observatory (PAO) in Argentina and the Telescope Array (TA) in Utah/USA. Covering an area of 3000 km² (PAO) and 700 km² (TA) with surface detectors, both detectors have collected many thousands of events with energies larger that $10^{19}$ eV and maximum energies of a few $10^{20}$ eV. TA consists of ~500 surface scintillation detectors (SD) with 1.2 km spacing and three stations of fluorescence detectors (FD) observing the night sky above the SD array [11,12]. INR is a member of the TA collaboration since 2008.

One of the main questions addressed by these arrays is the question about a possible cut-off of the cosmic ray spectrum and, if it exists, about the reasons for its existence. Is it that the accelerator cannot boost particles to higher energies, or are protons on the way through space absorbed by the so-called GZK effect? This effect has been proposed in 1966 by Kenneth Greisen (USA) and independently by Georgi Zatsepin and Vadim Kuzmin (both INR) [13,14]. The GZK limit is a theoretical upper limit on the energy of cosmic protons from intergalactic space. If their energy is above ~$5 \times 10^{19}$ eV, they will be absorbed by resonant interaction with the cosmic microwave background photons (CMB): $p + \gamma_{CMB} \rightarrow \Delta^+ \rightarrow p + \pi^0 \, / \, n + \pi^+$. As a result, protons with higher energy will not reach us or are energetically down-scattered if their sources are more than ~ 200 million light years away.

Both PAO and TA have confirmed the cut-off, with slightly different cut-off energies (somewhat higher for TA than for PAO). This discrepancy could be due to either different energy calibration (an effect which has been carefully studied by both collaborations) or with a source-dependent North/South discrepancy of the cut-off. TA found strong indications for a declination dependence of the effect in their own data, see Fig.5 [15]. The shown effect had a statistical significance of 4.0σ and – with more data – has meanwhile raised to 4.3σ [16].

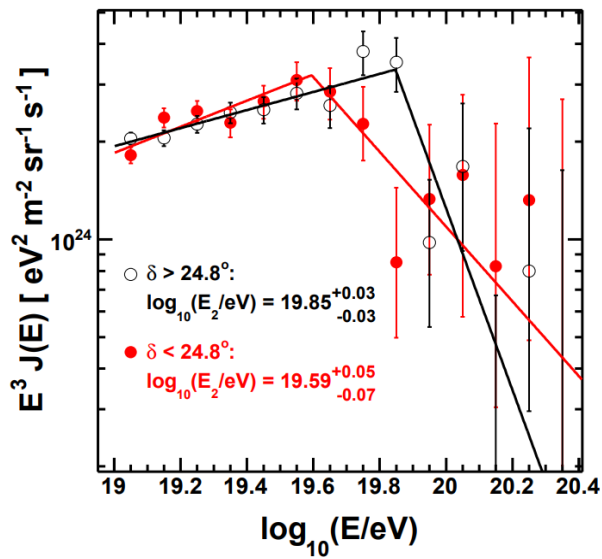

**Figure 5:** *TA Surface Detector spectrum for two declination bands using events with zenith angles from 0 to 55°. Superimposed are broken power law fit results for each of the declination bands [15].*

No clear answer could yet be given about the nature of the cut-off. A distinction between GZK-effect and maximum source power requires larger detectors and an improved ability to determine the mass composition.

TA has measured the cosmic ray energy spectrum over fantastic five decades, from a few $10^{15}$ eV to a few $10^{20}$ eV, using surface scintillation detectors, fluorescence telescopes and adding data from its low-energy extension TALE (see Fig.6). Presently it is being upgraded to almost the same size as PAO (TA×4), allowing a better determination of the cut-off energy, of the declination dependence of the spectra, and the possible confirmation of a "hot spot" at the cosmic ray sky map.

The INR group (initiated by Piotr Tinyakov, now Brussels) is strongly involved in the analysis of the TA data and interpretation of the results. The group members also participate in detector operation and maintenance. The hot spot for instance – the indication of an anisotropy at energies greater than 57 EeV – has been found by members of INR [17]. INR members have also developed a SD event reconstruction method which is employed to search for both ultrahigh-energy photons and neutrinos. They also study the mass composition with a multivariate machine-learning method. The present focus of the INR group is the development of the improved methods of composition analysis and photon search using convolutional neural networks.

### 2.3 Tunka-133

The Tunka air shower detector [18] addresses the energy range between the Baksan detectors and the Telescope Array, from ~$10^{15}$ eV up to a few $10^{18}$ eV. It is located in the Tunka valley, Siberia, and consists of a central array of 133 wide-angle air Cherenkov counters surrounded by six small satellite clusters with seven counters each. The central array has a diameter of 1 km, the small clusters are at distances of about 750 m. Each counter consists of a metallic cylinder, containing an upward-looking single PMT with 20 cm diameter. The small predecessor of Tunka-133 consisted of 25 stations containing the 37-cm photomultiplier ("QUASAR") which had been developed for the Baikal neutrino telescope (see 4.2). The lead institution is Moscow State University.

Tunka-133 is the first large-scale "Cherenkov timing detector" where the shower front is reconstructed from the arrival times at different space points, a technique which has been pioneered 30 years ago by a small array at the Canary Islands.

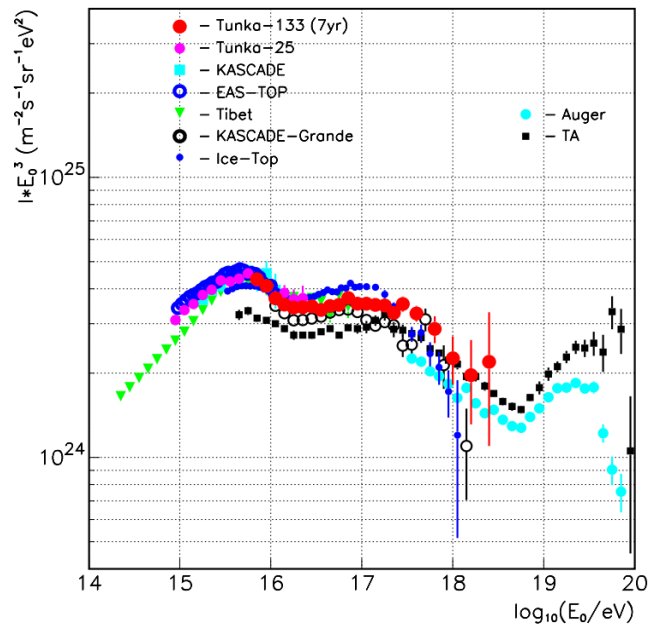

**Figure 6:** Comparison of energy spectra obtained at the Tunka site to other experimental results [18].

Figure 6 shows the energy spectrum of cosmic rays as measured with ground-based detectors, including data from the Telescope array, from Tunka-25 and from Tunka-133. The Tunka data are consistent with the spectra from KASCADE-Grande and IceTop (low energies) as well as those from the Telescope Array and the Pierre Auger Observatory (high energies). The timing technique which has been developed within the Tunka-133 project is meanwhile being used for gamma astronomy in the TAIGA project (see section 3). The figure also nicely demonstrates the apparent discrepancy between the cut-offs measured by TA and PAO which has been mentioned under 2.2.

*2.4 LHAASO*

The Large High-Altitude Air Shower Observatory (LHAASO) project [19] is a new generation instrument, being built at 4410 meters altitude in China, with the aim to study with unprecedented sensitivity energy spectrum, mass composition and anisotropy of cosmic rays in the energy range between $10^{13}$ and $10^{17}$ eV. At the same time, it will act as a wide aperture (~2 sr), 24 hours-per-day operating gamma ray telescope in the energy range between $10^{11}$ and $10^{15}$ eV (see section 3). LHAASO consists of the following major components (see Fig. 7): *a)* a 1 km$^2$ array (LHAASO-KM2A) for electromagnetic particle detectors (ED) divided into two parts: a central part including 4931 scintillator detectors 1 m² each in size (15 m spacing) to cover a circular area with a radius of 575 m and an outer guard-ring instrumented with 311 EDs (30 m spacing) up to a radius of 635 m, *b)* an overlapping 1 km² array of 1146 underground water Cherenkov tanks 36 m² each in size, with 30 m spacing, for muon detection (MD), *c)* A closely-packed surface water Cherenkov detector facility with a total area of about 78,000 m² (LHAASO-WCDA), *d)* 12 wide field-of-view air Cherenkov telescopes (LHAASO-WFCTA).

Due to the high altitude and the dense instrumentation, the sensitivity of LHAASO for cosmic rays reaches down to energies hundred times lower than those of other ground-based arrays (with the exception of the Tibet array, where the difference is a bit smaller). This will provide data widely overlapping with satellite and balloon experiments. The dense instrumentation with different types of detectors will also allow for a much better determination of the mass composition than that obtained with other ground-based cosmic-ray detectors. One interesting LHAASO component – not yet shown in Fig. 7 – is named ENDA (Electron-Neutron Detector Array). This is a novel technique which is of particular importance to distinguish hadron- from gamma-ray showers and to determine the mass composition. It has been developed over the last decade by the INR group of Yuri Stenkin.

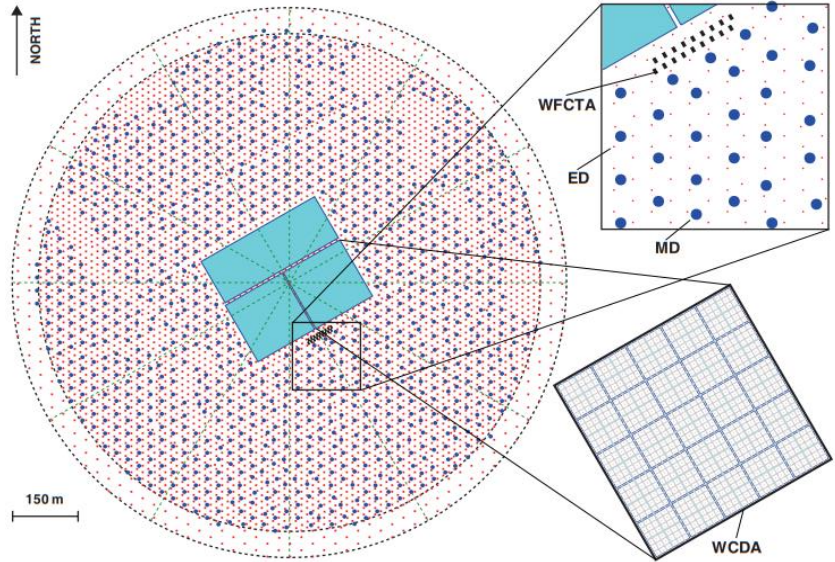

*Figure 7: The configuration of the LHAASO experiment.*

The ENDA detectors are of a similar type as the PRISMA detectors which have been operated above the ARGO detector (PRISMA-YBJ [20]) and are now operated around the NEVOD detector at MEPhI [21,22]. They respond to thermal neutrons, generated by the hadronic component of air showers. The original detectors were based on a granulated alloy of inorganic ZnS doped with Ag and $^6$Li-enriched LiF. $^6$Li-isotopes capture thermal neutrons via the reaction $^6$Li+n $\rightarrow$ $^3$H +$\alpha$+ 4.78 MeV. The light yield is ~160, 000 photons per neutron capture, slow neutrons will be captured with an efficiency of 20%. The structure of a PRISMA-detector is shown in Fig. 8. In its final stage, ENDA is conceived to consist of a total of 400 stations, of which 16 are already in operation).

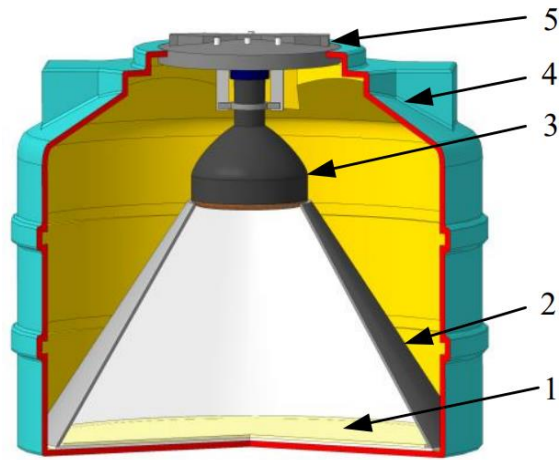

*Figure 8: Structure of a PRISMA detector. The ENDA detectors for LHAASO are very similar in design. 1: ZnS(Ag)+$^6$LiF scintillator (replaced by an alloy of ZnS and boron composition for ENDA); 2:light reflecting cone; 3: PMT FEU-200 (for ENDA it is a 4´PMT Hamamatsu CR165); 4: tank wall; 5: lid. [21].*

LHAASO marks a huge leap both for cosmic-ray and gamma-ray physics. I will address first exciting results on gamma-astronomy in section 3.

## 3. Gamma-ray astronomy

### 3.1 From the Crimea telescope to modern Imaging Air Cherenkov Telescopes

Among the three ways to explore the Universe at highest energies (cosmic-ray, gamma-ray and neutrino approach), gamma-ray astronomy is the most successful strategy so far. With almost 200 known sources in 2020, it has revealed a very detailed picture of the high-energy cosmos. The key method is the imaging of cosmic air showers via the Cherenkov light generated in the atmosphere (IACT, for **I**maging **A**ir shower **C**herenkov **T**elescopes).

Cherenkov light flashes from air showers have been first detected by Bill Galbraith and John Jelley in 1952 [23]. Among those who interpreted the flashes as caused by air showers was Aleksandr Chudakov [24]. He has also been the first to realize the idea of a calorimetric measurement of air shower cascades with the help of the emitted Cherenkov light [25]. His experiment in the Pamir Mountains pioneered the detailed study of Cherenkov radiation emitted by air showers.

In 1961, Georgi Zatsepin proposed to Chudakov (who was just installing an air shower Cherenkov detector in the Crimea) to employ his array to search for high-energy sources of gamma rays. Both published a common paper on that idea [26]. In the following years, Chudakov took the worldwide first step into high-energy gamma astronomy by operating his array as a gamma-ray telescope and observing several candidate sources. The telescope consisted of 12 mirrors of 1.5 m diameter, each focusing the light to a single photomultiplier. The observed sources included Cygnus-A and the Crab Nebula but, in the absence of a signal, Chudakov only could derive upper limits on the gamma-ray flux from these sources. Seen from today, this is no surprise: compared to the cosmic-ray background, the gamma-ray fluxes are much too small to be identified without using either imaging or timing techniques [27].

It took nearly three decades until a breakthrough was achieved. In 1989, the Whipple telescope in Arizona identified the first TeV gamma-ray source, the Crab nebula [28]. The telescope consisted of a 10-meter fragmented mirror focussing its light to a mosaic of photomultipliers. The pattern of the fired photomultipliers imaged the shower. Cuts on the shower parameters allowed distinguishing a gamma-ray signal from the background of cosmic-ray initiated showers.

Until 1996, Whipple had identified two additional sources (the Active Galaxies Markarian 421 and 502). In 2005 the number of sources had increased to ~80 and in 2015 to ~150. Most of the progress of the last 1½ decades has been achieved by three multi-telescope arrays, allowing for stereo imaging: H.E.S.S. in Namibia (5 telescopes), MAGIC on La Palma (2 telescopes) and VERITAS in Arizona (4 telescopes). Meanwhile, the communities of these projects have united to build the huge Cherenkov Telescope Array (CTA), with two sites, the one in La Palma, the other in Chile.

Russian institutes have been neither partners of H.E.S.S., MAGIC or VERITAS, nor are they partners in CTA. But INR is partner of two gamma-ray/cosmic-ray observatories which – among other techniques – make also use of the imaging technique: TAIGA in Siberia and LHAASO in China.

### 3.2 TAIGA

TAIGA stands for **T**unka **A**dvanced **I**nstrument for cosmic ray physics and **G**amma **A**stronomy. It is a follow-up project of the Tunka-133 array described in section 2. When finished, it will consist of 120 wide angle timing detectors (named HiSCORE [29]), at least three imaging telescopes (IACTs) and a large number of muon counters [30, 31]. HiSCORE follows the same detection principle as Tunka-133, but with a much better time resolution, i.e. directional precision. It will allow good γ/hadron separation at high energies. The imaging telescope will support γ/hadron separation at low energies and yield a superior directional resolution. The muon counters will also support γ/hadron separation. Figure 9 shows a view of a timing detector and an imaging telescope. The TAIGA collaboration includes a number of Russian Institutes (with MSU as lead institution) and three German partners (DESY/Humboldt Univ., Univ. Hamburg and MPI Munich).

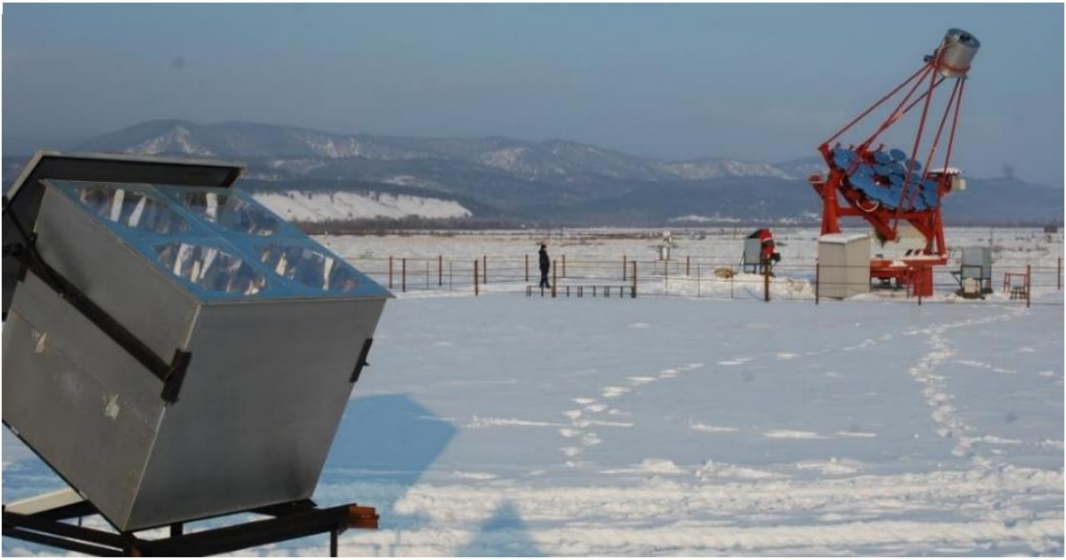

*Figure 9:* *View of a HiSCORE timing detector and an Imaging Telescope of the TAIGA array.*

With the stereoscopic operation of the first two IACTs, the identification of the Crab Nebula up to ~50 TeV energy has been achieved [31] (adding the HiSCORE information even up to energies up to ~100 TeV). While observation of the Crab Nebula is just a first test of proper operation, observation of other sources is underway, and preliminary results indicate gamma-ray signals from two of them [32].

*3.3 LHAASO*

The central part of LHAASO covers a 78.000 m² area which is closely packed with water tanks (LHAASO-WCDA). Gamma-ray astronomy with large water tanks has been already very successfully practised with the HAWC array in Mexico, located at a similar altitude as LHAASO but with a more than two times smaller area. The huge leap with LHAASO, however, is given by the multitude of other detectors which are able to discriminate between hadron and γ-showers and which cover an area ~16 times larger than that of WCDA alone.

Actually, the first gamma-ray results have been obtained without WCDA, using only 50% of the KM2A component already installed in 2019. KM2A consists of surface scintillation detectors (ED) and underground water Cherenkov tanks (MD), with 15 and 30 meters spacing, respectively. ED serves for detection of the electromagnetic part of the shower, MD for the detection of muons. In [33], the collaboration presents the results of an analysis based on the first five months of data from the KM2A half-array to study the Crab Nebula as standard candle at very high energy. The statistical significance of the observed gamma-ray signal is 19.2σ at 10-25 TeV, 28.0σ at 25-100 TeV and 14.7σ at >100 TeV, and the gamma-ray angular distributions around the source are fairly consistent with the point spread function obtained by simulations in the corresponding energy intervals (0.46°, 0.29° and 0.16° respectively).

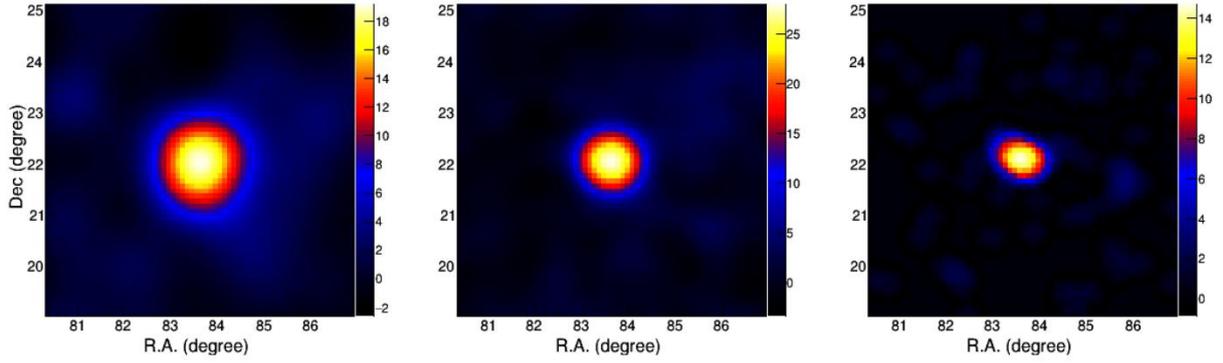

*Figure 10: Significance maps centred on the Crab Nebula at three energy ranges: left 10-25 TeV, middle 25-100 TeV, right > 100 TeV. $\sigma_S$ is the standard deviation of the 2-dimension Gaussian taken according to the point spread function of KM2A. The colour represents the significance. S is the maximum value in the map (figure taken from [33]).*

While the present paper is being written, also results from the first of the 25 WCDA sections have been released [34], again focusing on the Crab Nebula. Figure 11 reproduces the gamma-ray spectrum for the Crab as obtained by WCDA-1, together with the spectrum deduced from 50% of KM2A, both compared to results from other detectors. At low energies, the WCDA-1 results are in good agreement with other data, while above 10 TeV one observes perfect agreement of the KM2A data with HAWC results.

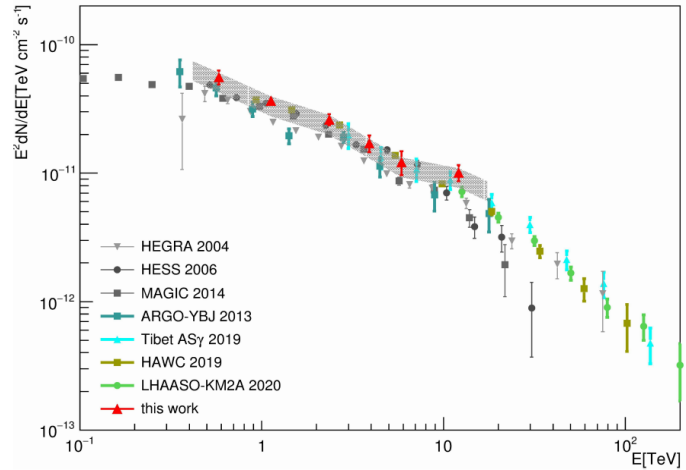

*Figure 11: The spectral energy distribution of the Crab Nebula in the energy range from 800 GeV to 13 TeV measured by WCDA-1 (red filled triangles) and 50% of KM2A above 10 TeV (green filled circles). The error bars are statistic errors while the shaded area represents the systematic error. Measurements of other experiments are plotted for comparison [34].*

Both WCDA and KM2A have a wide-angle acceptance and, differently to IACTs, can be operated 24 hours per day. Needless to say, that the large field of view has led to the detection of more sources than only the Crab Nebula, several of them not yet seen by IACTs and several of them with energies well beyond 100 TeV. These results have not yet been published but are being shown on conferences.

The initial LHAASO results herald a new era in gamma-astronomy. Whatever the involvement of INR in the analysis of the gamma-ray data is, it appears obvious that in the spirit of multi-messenger astronomy LHAASO will be an important partner for the neutrino telescope Baikal-GVD (section 4).

## 3.4 Carpet as a gamma-ray detector

The initial focus of the Carpet array (see section 2) has been on cosmic-rays. But it has also an interesting history in gamma-ray research reaching back to the 1980s. The following recapitulation of this history follows a paper by Alexander Lidvansky [35].

In 1980s, considerable efforts on ultra-high energy gamma ray astronomy were stimulated by a paper of a group from Kiel (Germany). They operated a scintillator array and claimed a signal from the X-ray source Cygnus X-3 in the multi-TeV range. It is an irony of history that an obviously erroneous observation (as it turned out later) did considerably stimulate gamma-ray astronomy. This included also searches with the Carpet array. Although no clear confirmation of the Cygnus results was obtained, neither by the Kiel group itself nor by many other experiments, several indications of a positive effect kept the campaign alive over several years. Carpet obtained just an upper limit on the *steady* gamma-ray flux from Cygnus X-3 at $10^{13}$ eV, but detected an apparent three-day *flare* from the Cygnus direction [36]. This flare (October 14-16, 1985) had appeared some days after an exceptionally strong radio outburst from Cygnus X-3. Shortly later, a gamma-ray telescope in India announced signals from Cygnus X-3 on October 10 -12. In [37] Veniamin Berezinsky proposed a model explaining the considerable delay between the radio signal and VHE and UHE gamma signals for this event, but taken altogether and judged from today, the situation stayed unresolved.

In 1989, the Carpet group announced a possible burst from the Crab Nebula at an energy of ~100 TeV [38]. This flare from February 23, 1989 was confirmed by the Kolar Gold Field (KGF) group in India. Two other groups (Tien Shan/Soviet Union and EAS-TOP/Italy) also reported some indications of an excess. In [35], Alexander Lidvansky summarizes the situation:

> *"It is understandable that [these] results were met with some distrust: at that time a burst of so high energy (hundreds of TeV) seemed too exotic, especially from the source, which, though it had just been recorded by Cherenkov telescopes at TeV energies, served as a pattern of stability (so that for a long time it was considered as "standard candle" in very high energy astronomy and in X-ray astronomy). However, many years later the flares from the Crab nebula were detected by satellite gamma-ray telescopes AGILE and Fermi-LAT at energies of a few hundred MeV. ... One of the AGILE flares had a time structure fully similar to that of the burst on February 23, 1989 (accounting for a single scale factor). This fact compels one to take the latter most seriously: after all, this signal from a celestial source has been recorded at a highest energy ever detected."*

More than two decades later, in 2013, IceCube announced the discovery of a diffuse flux of cosmic neutrinos at > 30 TeV and thereby triggered a new perspective for Carpet: the search for a diffuse flux of high-energy gamma-rays. The gamma rays should be co-produced with the neutrinos, the neutrinos stemming from the decay of charged pions and the gamma-rays from the decay of neutral pions.

To improve the gamma-hadron separation, the area of Carpet's Muon Detector (MD) is going to be increased from 175 m² ("Carpet-2") to 410 m², and in a second step to 615 m². The surface array will be increased by additional detectors, each of them containing nine scintillation counters of 1 m² area. By now, the first upgrade ("Carpet-3" with MD ~ 410 m²) has been realized and the commissioning of the MD is underway. Figure 12 shows a view of the tunnel with the muon detectors.

The Carpet collaboration has recently released several papers on gamma-ray searches with Carpet-2, for instance a search for PeV gamma-rays associated with IceCube high-energy neutrino events [39], a search for a flux of diffuse gamma rays with energy above 700 TeV [40] or a search for gamma rays above 100 TeV in coincidence with HAWC and IceCube alerts [41]. All these searches resulted in upper limits only. Figure 13 compares the effective areas of IceCube and Carpet-2 for gamma rays averaged over the Northern hemisphere and demonstrates that the Carpet-2 area for vertical tracks is comparable to the IceCube effective area for neutrinos.

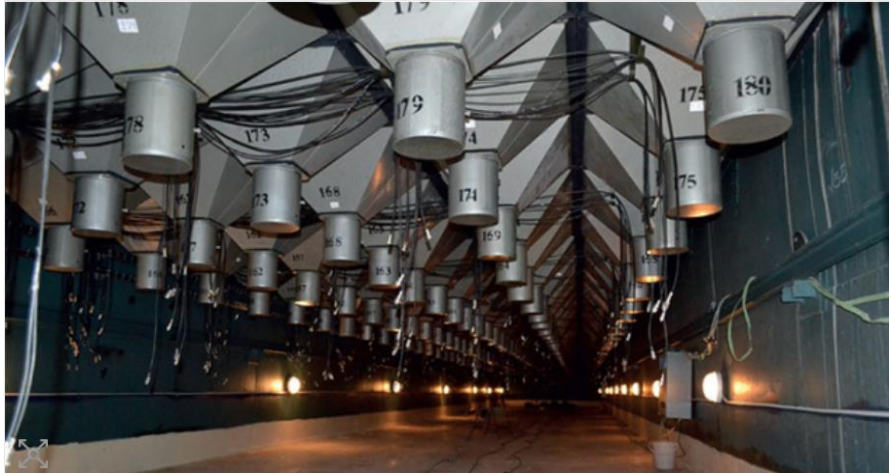

***Figure 12:*** *Inside the muon detector of Carpet*

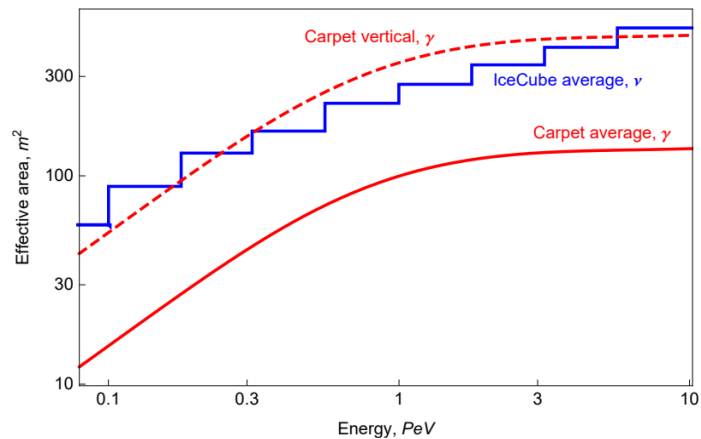

***Figure 13****: Comparison of effective areas of Carpet–2 (red continuous lines, photon detection, present analysis) and IceCube (blue step line, neutrino detection, muon tracks, average over the Northern hemisphere). For Carpet–2, the full line gives the average over the field of view while the dashed line corresponds to vertical events. Figure taken from [39].*

## 3.5 LHAASO vs. Carpet and TAIGA

The fantastic sensitivity of LHAASO dwarfs the sensitivities of TAIGA and Carpet. It is hard to judge, to what extent these two detectors will be able to compete with LHAASO and how they can profit from potential niches. One example are transient events with the source outside LHAASO's field of view due its longitude. Another example is the coverage of the parts of the Northern hemisphere which are not in the field of view of LHAASO. For instance, TAIGA is ~23° further North than LHAASO. The gamma-ray source in the Tycho Supernova Remnant, almost inaccessible for LHAASO, will be in TAIGA's field of view during 500 hours per year.

Whatever the answer will be: INR is part of a voyage into promising territories.

## 4. Neutrino astronomy

### 4.1 Early history

Not long after the discovery of neutrinos by Frederick Reines and Clyde Cowan in 1956, first ideas came up on how to detect atmospheric and cosmic neutrinos. The first who looked in some detail to this possibility were two later members of INR, namely Moisej Markov and his student Igor Zheleznykh. In his diploma work from 1958, Zheleznykh investigated the possibility to detect atmospheric neutrinos underground. In the 1960 *Annual Review of Nuclear Science*, also Kenneth Greisen and Reines discussed the motivations and prospects for such detectors [42,43]. In the same year, on the 1960 *Rochester Conference*, Markov published his ground-breaking idea [44] *"...to install detectors deep in a lake or in the ocean and to determine the direction of charged particles with the help of Cherenkov radiation."*[2] Going to open water appeared to be the only way to reach detector volumes beyond the scale of $10^4$ tons.

From the perspective of the 1960s and early 1970s, the study of atmospheric neutrinos appeared equally interesting as the search for cosmic neutrinos. For instance, with atmospheric neutrinos one hoped to solve the question of four-fermion interactions vs. Yukawa-type interaction by studying the energy-dependence of the neutrino cross section. First reliable calculations of the flux of atmospheric neutrinos were published in 1961 by Markov and Zheleznykh [45] and by Zatsepin and Kuzmin [46], all of them later members of INR. In 1965, atmospheric neutrinos were discovered in deep mines in South Africa and in India. In the following decades, deep underground detectors would play a triumphal role for studying atmospheric neutrinos and discovering neutrinos from the Sun and from the Supernova SN1987 A.

In this concert, BUST in the Baksan Laboratory was one of the pioneers. In the 1970s it became clear that a reasonable W-mass would be too high to lead to an observable flattening of the neutrino cross section in the GeV region (to which BUST had to focus, due the fact that the flux falls steeply with energy and the area of BUST is not very large). On the other hand, the hypothesis of *neutrino oscillation* found increasing acceptance. For sufficiently large mixing angles and appropriate mass differences, oscillation should lead to a deficit of atmospheric neutrinos which had passed the Earth. Actually, BUST was the first to constrain the oscillation parameters with the first years of collected data [47]. Also in 1981, the BUST collaboration published a world-best limit on the proton lifetime (which, however, kept the world-record for only half a year [48]). From the early 1980s until the late 1990s, BUST also delivered world-leading limits on super-heavy magnetic monopoles in the velocity range of $10^{-3}$ to $10^{-1}$ c [49], before it was bypassed by MACRO in the Gran Sasso Laboratory in Italy. BUST also provided important limits on the interaction cross section of WIMPS (Weakly Interacting Massive Particles) which might have accumulated in the center of the Earth, annihilate with each other and lead to an excess of vertically upward moving neutrinos [50].

For the detection of high-energy neutrinos from cosmic accelerators, however, underground detectors turned out to be too small. That put Markov's proposal back to the agenda, and in 1973 first discussions came up to build a large underwater detector.

These discussions led to the legendary DUMAND project. From the USSR, Georgy Zatsepin was among its founding fathers. The 1978 design of the first configuration consisted of 22,698 optical modules and covered a volume of 1.26 km³, following the idea that "*...the size of the array was based on relatively scant information on the expected neutrino intensities and was difficult to justify in detail*". However: "*The general idea was that neutrino cross sections are small and high-energy neutrinos are scarce, so the detector had better be large*." (cited from [51]). DUMAND was planned to be deployed off the coast of Big Island/Hawaii.

Soviet participation in the DUMAND project was strong from the beginning and represented by names like Aleksandr Chudakov, Veniamin Berezinsky, Leonid Bezrukov, Boris Dolgoshein, Anatoly Petrukhin and Igor Zheleznykh (the first three and the last as members of INR). In 1979, one of the DUMAND Workshops was held in Khabarovsk and at Lake Baikal. However, in the context of the Soviet intervention in Afghanistan, in 1980 the Reagan administration terminated the cooperation. Although Chudakov (together with American, Japanese and European partners) still signed an open letter from July 26, 1980 which declares the formation of a DUMAND collaboration board, it was clear that Russia had to find its own way.

While from now on the INR activities on underwater detectors were focused to Lake Baikal (see next subsection), a small group around Igor Zheleznykh started Ocean explorations for a "Soviet DUMAND". In 1989, they cruised with the Soviet vessel "Mendeleyev" in the Mediterranean Sea in order to determine a suitable place for a neutrino telescope. The prototype of an optical module with four large photomultipliers carried by a titanium frame was tested, and the depth-dependence of the muon flux was measured down to 3500 meters [52]. Two years later, together with Greek collaborators, a prototype of a NESTOR "floor" with ten PMTs was tested from the Vityaz Vessel. But, with the successful operations at Lake Baikal and with the technical and financial problems of the NESTOR project, these activities were terminated in the following.

## 4.2 The Baikal neutrino telescope: from site studies to NT200

Early 1980, Aleksandr Chudakov proposed to use the deep water of Lake Baikal in Siberia as a test site for a Soviet DUMAND. The advantages of Lake Baikal seemed obvious: it is the deepest freshwater lake on Earth, with its largest depth at nearly 1700 meter, it is famous for its clean and transparent water, and in late Winter it is covered by a thick ice layer which allows installing winches and other heavy technique and deploying underwater equipment without any use of ships.

On October 1, 1980, following a proposal of Moisey Markov, the *Laboratory of Neutrino Astrophysics at High Energies* as a department of INR was founded. The group was led by Grigory Domogatsky, a theoretician, flanked by Leonid Bezrukov as leading experimentalist.

In 1981, first shallow-depth experiments with small PMTs were started. A site in the Southern part of Lake Baikal, at a distance of 3.6 km to shore and at a depth of about 1370 m was identified as the optimal location for a detector which would be installed at a depth of about 1.0-1.1 km. Components could be deployed in a period between late February and early April from the ice cover, and operated over the full year via a cable to shore.

After operation of first underwater modules with a 15-cm PMT in 1982, in the following year a small string was operated for several days. In 1984, a first *stationary* string was deployed (Girlanda-84) and recorded downward moving muons [53]. It consisted of three floors ("sviaskas") each with four PMTs in two pressure-tolerant cylinders of glass-fibre enforced epoxy. At that time, no pressure tight glass spheres where available in the USSR. The ends of the cylinders were closed by caps of plexiglass (see

Fig. 14, left). The PMT FEU-49 was the same type as the PMTs of BUST and CARPET, with a 15 cm flat photocathode and modest amplitude and time resolution.

The 1984 string was followed by another stationary string in 1986 (Girlanda-86). Data from this string were used to search for slowly moving bright particles like magnetic monopoles [54]. Typical GUT versions predict monopoles with masses $10^{16}$ GeV and more. These monopoles would have typical velocities of $v/c = 10^{-4} - 10^{-3}$. They might catalyse baryon decays along their path (Rubakov-Callan effect [55]) which could be detected via the Cherenkov light from the decay particles. Girlanda-86 consisted of six floors at very large vertical spacing of 50 m. It was optimized to detect tracks of non-relativistic particles much brighter than muon tracks. Data from Girlanda-86 set stringent limits not only on the flux of magnetic monopoles with high catalysis cross section, but also on the flux of Q-balls (super-heavy hypothetical soliton solutions of field theories which would generate a similar light pattern as monopoles) [56].

Girlanda-84 took data for a total of 50 days and then sank down due to leaking buoys which held the string in vertical position. The cable penetrators through the epoxy cylinders as well as the cap-to-cylinder hermetic connection also tended to leak and were a notorious source of headaches. Moreover, it was clear that the PMT used was much too small and too slow for a neutrino telescope. Therefore, a technology using glass spheres and a new type of photo-sensor were developed. Leonid Bezrukov started the development of an equivalent Russian device, the QUASAR, in cooperation with the EKRAN company in Novosibirsk. The QUASAR (Fig. 14, right) is a hybrid device similar to the PHILIPS 2600 developed for the DUMAND project. Photoelectrons from a 370 mm diameter cathode are accelerated by 25 kV to a fast, high-gain scintillator near the centre of the glass bulb. The light from the scintillator is read out by a small conventional photomultiplier. The QUASAR had an excellent 1-PE resolution, clear distinction between 1-PE and 2-PE pulses, a time jitter as small as 2 ns and negligible sensitivity to the Earth's magnetic field [57].

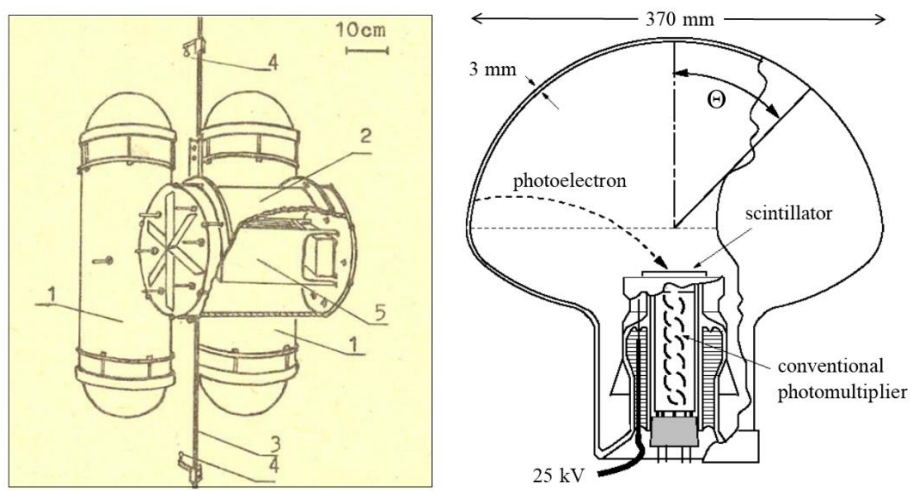

**Figure 14:** *Left: A "svjaska" as used in GIRLANDA-86, with 4 PMTs in 2 cylinders (1) covered by hemispheric plexiglass caps. The "backpack" module (2) contains electronics (5). Calibration LEDs (4) are fixed at the geophysical cable (3). Right: The QUASAR-370 phototube.*

In 1987, the Baikal experiment was approved as a long-term direction of research by the Soviet Academy of Sciences and the USSR government which included considerable funding. A full-scale detector (yet without clear definition of its size) was planned to be built in steps of intermediate detectors of growing size. One year later, my own group from the East German Institute of High

Energy Physics in Zeuthen joined the Baikal experiment. After German unification in 1990, the institute became part of DESY. Immediately we had access to the Western market and could contribute with German glass spheres and some underwater connectors to the strings which were deployed in 1991 to 1993. In parallel, Russian spheres were developed in collaboration with industry, as well as penetrators and connectors which tolerated water depths down to 2 km – not suitable for large-depth Ocean experiments but sufficient for Lake Baikal.

In 1989, a preliminary version of what later was called the NT200 project was developed, an array comprising approximately 200 optical modules. The final version of the project description was finished in 1992 [58]. At this time, the participating groups came from INR Moscow, Univ. Irkutsk, Moscow State Univ., Marine Techn. Univ. St. Petersburg, the Polytechnical Institutes in Niszhni Novgorod and Tomsk, JINR Dubna, Kurchatov Inst. (Moscow), Limnological Inst. Irkutsk (all Russia), DESY-Zeuthen (Germany) and KFKI Budapest (Hungary).

NT200 (Fig. 15, left) was an array of 192 optical modules carried by eight strings which were attached to an umbrella-like frame consisting of 7 arms, each 21.5 m in length. It spanned 72 m in height and 43 m in diameter. A stunning feature of NT200 was the finely balanced mechanics of this frame, with all its buoys, anchor weights and pivoted arms. It was designed by Andrei Panfilov who sadly passed away in 2019 during a winter expedition. The optical modules with the QUASAR-370 phototubes were grouped pair-wise along a string. In order to suppress accidental hits from dark noise and bio-luminescence, the two photomultipliers of each pair were switched in coincidence. The time calibration was done using several nitrogen lasers in pressure-tight glass cylinders.

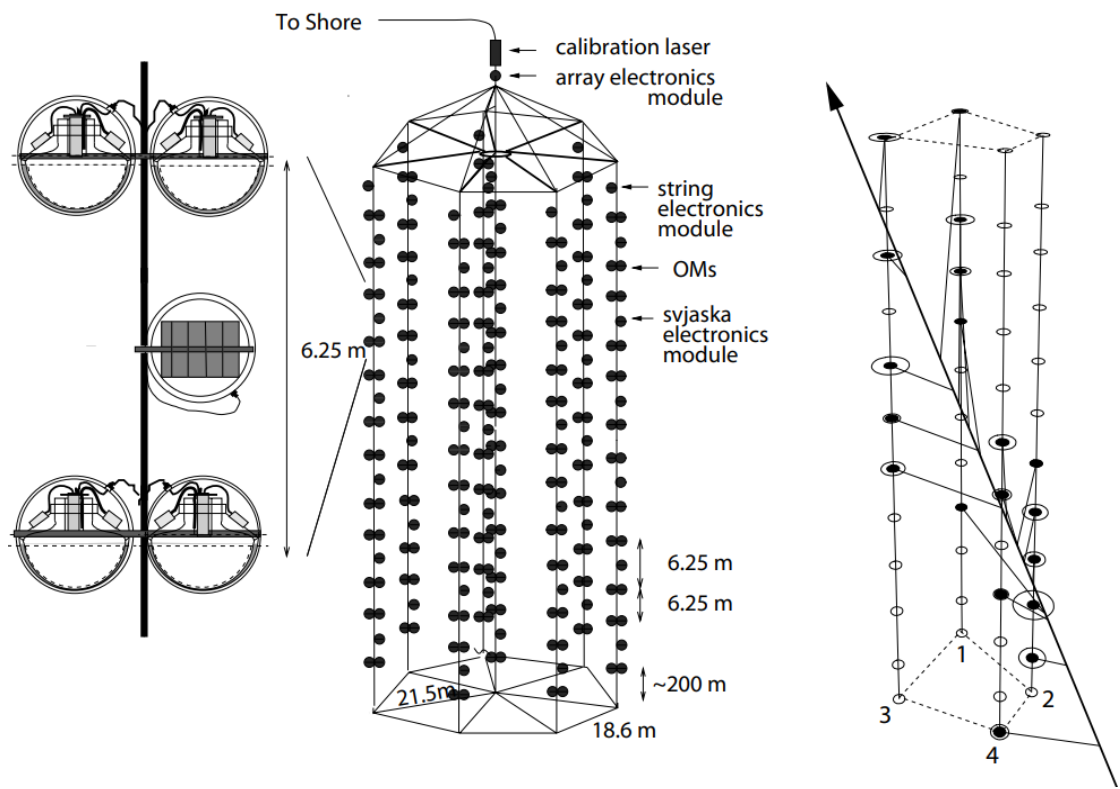

**Figure 15:** *Left: The Baikal Neutrino Telescope NT200. Right: One of the first upward moving muons from a neutrino interaction recorded with the 4-string stage of the detector in 1996 [60]. The Cherenkov light from the muon is recorded by 19 channels.*

The construction of NT200 coincided with the decay of the USSR and an economically desperate period. Members of the collaboration and even some industrial suppliers had to be supported by grants from Germany; nevertheless, many highly qualified experimentalists left the collaboration and tried to survive in the private sector. Over a period of three years, a large part of the food for the winter campaigns at Lake Baikal had to be bought in Germany and transported to Siberia. Still, a nucleus of dedicated Russian physicists heroically continued to work for the project.

Under these circumstances the construction of NT200 extended over more than five years. It started with the deployment of a 3-string array with 36 optical modules in March/April 1993. The first two upward moving muons, i.e. neutrino candidates, were separated from the 1994 data (see Fig. 16) and Grigory Domogatsky sent the following telefax to me:

*Дорогой Кристиан, сначала самое интересное: При анализе данных 1994ого года одно событие категорически сопротивляется всем тестам, и кажется это нейтрино. Естественно, я не хотел бы решать, декларируем ли мы это событие как первое ясное нейтрино в подводном детекторе. Но это можно называть очень на нейтрино похожим событием, и мне кажется, что очень желательно сообщить об этом через 2 или 3 недели достаточно широкой общественности.*

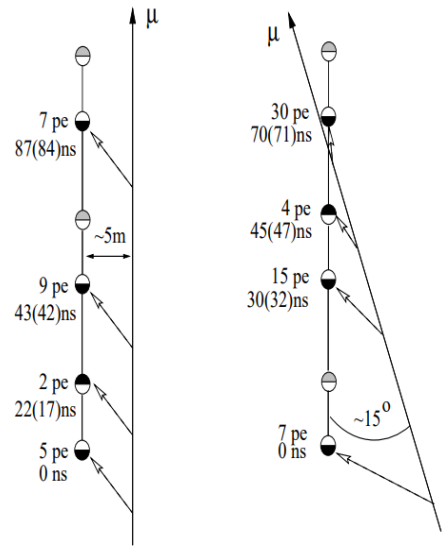

Figure 16: *The two neutrino candidates from NT-36. The hit PMT pairs (channels) are marked by black. Numbers give the measured amplitudes (in photoelectrons) and times with respect to the first hit channel. Times in brackets are those expected for a vertical going upward muon (left) and an upward muon passing the string under 15° (right) [59].*

NT-36, as small as its dimensions were, was the first underwater array capable of three-dimensional reconstruction of muon tracks and the first to separate candidates for upward moving muons from neutrino interactions. In 1996, a 96-OM array with four NT200 strings was operated [60] and it provided the first very clear "textbook" neutrinos like the one shown in Fig. 15 (right).

NT200 was completed in April 1998 [61] and has taken data until 2008. The basic components have been designed and built in Russia, notably the mechanics of the detector, the optical module, the underwater electronics and the cabling. All non-Russian contributions came from DESY: the laser time calibration system, a transputer farm in the shore station for fast data processing, an online monitoring system and a special underwater trigger tailored to register slowly moving very bright particles (as GUT monopoles).

The small spacing of modules in NT200 led to a rather low energy threshold of about 15 GeV for muon detection. About 400 upward muon events were collected over ~5 years. This comparatively low number reflects the notoriously large number of failures of individual channels during a year. Still, NT200 could compete with the much larger AMANDA for a while [3] by searching for high energy cascades below NT200, surveying a volume about ten times as large as NT200 itself [62,63] (see Fig.17). In order to improve pattern recognition for these studies, NT200 was fenced in 2005 and 2006 by three sparsely instrumented outer strings (6 optical module pairs per string). This configuration was named NT200+, but suffered from several problems (from the new strings as well as from the meanwhile antiquated NT200 itself), so that no satisfying statistics and no convincing results have been obtained. On the other hand, it served as a testbed for some concepts of data transmission which later were applied in Baikal-GVD.

NT-200 has set upper limits on the flux of neutrinos from dark matter annihilation in the centre of the Earth and in the Sun, on the flux of relativistic monopoles [64] and the flux of neutrinos from Gamma Ray Bursts in the Southern hemisphere. The most competitive limits have been those obtained for the diffuse flux of cosmic neutrinos.

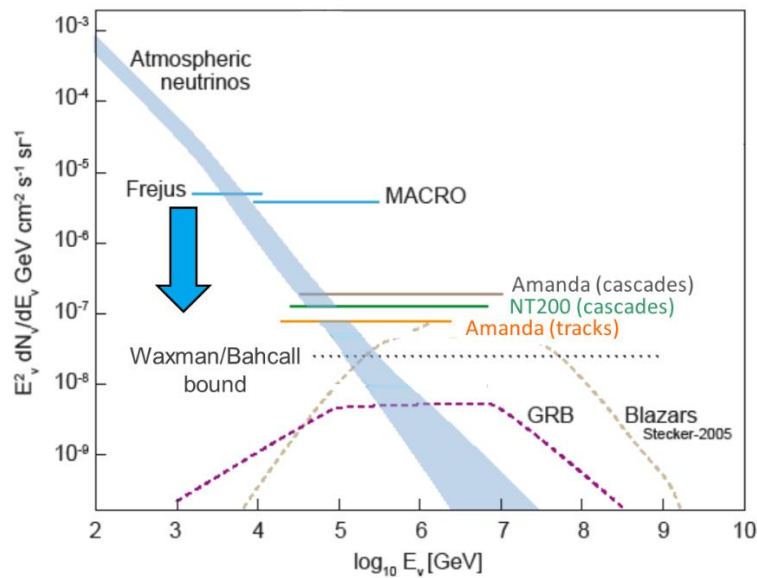

*Figure 17: Experimental limits on the diffuse flux of neutrinos as obtained in 2003, compared to theoretical upper bounds and model predictions. See text for details.*

Figure 17 shows the status of upper limits on the diffuse flux of neutrinos in 2003. At this point, AMANDA and NT200 had made already a huge leap by ~1½ orders of magnitude compared to underground detectors (FREJUS and MACRO). Limits were coming close to very optimistic model predictions as the one shown for blazars (i.e. Active Galactic Nuclei with their jet pointing to us).

---

[3] The geometric volume of NT200 was only ~10⁻⁴ km³, compared to 0.015 km³ for AMANDA.

They were just bypassing a theoretical upper bound which Veniamin Berezinsky had derived from low-energy gamma-ray data (not shown in the figure), but were still a factor of 2 to 3 above the Waxman-Bahcall bound which had been derived from cosmic ray data. As mentioned above, NT200 could compete with early AMANDA results, due to the possibility to look beyond its geometric volume (a possibility which is given only in water, due to the smaller light scattering compared to ice). A few years later ANTARES in the Mediterranean Sea joined the race, moreover first IceCube data appeared, so that at the end of the decade NT-200 limits became less important.

### 4.3 Opening the window: IceCube

The IceCube detector can be considered as the realization of the initial DUMAND plan: a detector of one cubic kilometer volume. IceCube consists of 86 strings, each with 60 optical modules, which are installed in the 3 km thick ice cover of the geographical South Pole. IceCube's predecessor was named AMANDA (**A**ntartic **M**uon **A**nd **N**eutrino **D**etection **A**rray), a detector 0.5 km high, with a diameter of 200 m. The DESY group joined AMANDA in 1994, keeping however actively its membership in the Baikal experiment. Our rationale for this decision was the unpredictability of the situation in Russia, both politically and economically, so we wanted to have a second mainstay. AMANDA was installed between 1995 and 2000 and took data until 2008. With the appearance of the Waxman-Bahcall bound (established in 1998) however, it had become clear that the discovery potential of a detector with only 0.015 km³ volume was almost zero and that AMANDA could be considered only a prototype for a larger array. This detector – IceCube – was constructed between 2004 and 2010.

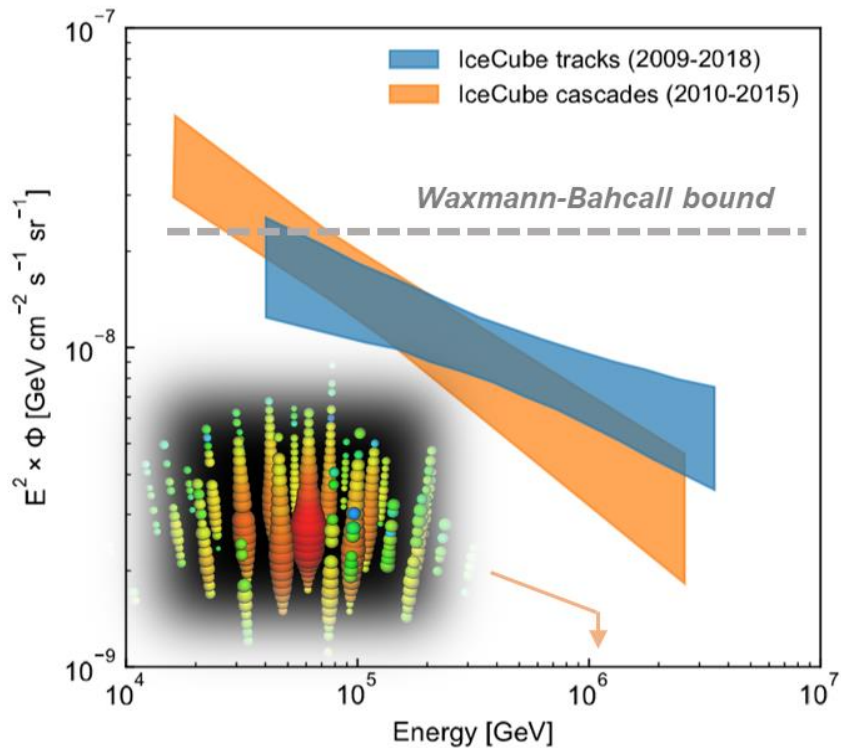

**Figure 18:** *Flux of cosmic neutrinos multiplied with the squared energy, as a function of energy. The dependencies with their error corridors are shown separately for track-like and cascade-like events. Bottom-left, a display of one of the two first PeV-events is shown. Each blob stands for a hit photomultiplier, the size indicates the signal amplitude, the colour the time (red: early, yellow: late).*

It took until 2012 when the first two events with a high probability for an astrophysical origin were found, both with an energy of about one PeV. Both were cascade-like events. One of them, with a reconstructed energy of 1.14 PeV, is shown in Fig. 18. With an improved analysis (again focusing to cascade-like events), about twenty additional cosmic neutrino candidates could be separated [65]. This was the long hoped-for discovery of high-energy cosmic neutrinos! Meanwhile, this number has increased to almost hundred. Moreover, a similar number of track-like events of likely cosmic origin have been found. Figure 18 (modified from [66] shows the energy spectra derived for cascade-like and track-like events.

The flux at 100 TeV almost saturates the Waxman-Bahcall bound. The slope is much steeper than the conventional $E^{-2}$ dependence suggested by Fermi acceleration. The different slopes for cascade- and track-like events are not yet explained and call for independent data from other detectors.

Up to now, IceCube has strong indications for a few point sources, but none of them can already be considered a "crystal-clear" discovery. The detector is located at the South Pole and – taking the Earth as filter – predominantly observes the Northern hemisphere. This situation calls for detectors on the Northern hemisphere with a better view to the Southern sky and the central parts of the Galaxy. Moreover, the search for point sources requires detectors with better angular resolution (i.e. in water, not ice). And last not least, to clarify the puzzle of the different slopes of events of different topology, one needs experiments with complementary systematics. Two such detectors are presently under construction: GVD in Lake Baikal and KM3NeT in the Mediterranean Sea.

### 4.4 The Baikal neutrino telescope: The GVD project

While in 2008 NT200 had exceeded its reasonable operating life, the Baikal collaboration envisaged a detector on the cubic kilometre scale, which initially was named NT1000 and later renamed GVD (for **G**igaton **V**olume **D**etector). GVD was conceived to consist of independent modules, the so-called "clusters", each comprising eight strings. The QUASAR phototube was discarded in favour of a much easier-to-handle, conventional 10-inch photomultiplier from Hamamatsu. Long-term in-situ tests with short strings were performed in 2008 and the following years [67]. In 2011 a small prototype cluster was deployed. Year after year, weak points were identified and fixed, so that in 2016 a first almost full-scale cluster was deployed. Acknowledging the extraordinary strong role of JINR Dubna, this cluster was christened "DUBNA".

Each GVD string holds 36 optical modules housing a downward pointing 10-inch PMT plus electronics modules and various sensors. The OMs are installed with a 15 m vertical spacing, resulting in 525 m instrumented string length. The strings form a heptagon of ~120 m diameter, with one central string and seven peripheral strings. The clusters are also arranged in a hexagonal pattern, with a ~300 m distance between the cluster centers. Additional strings, solely equipped with high-power pulsed lasers, are installed in-between the clusters. They are used for detector calibration and light propagation studies. The detector layout is optimized for the measurement of astrophysical neutrinos in the TeV–PeV energy range.

The first cluster of Baikal-GVD was deployed in 2016. Both in 2017 and 2018, one new cluster was added, followed by two more in 2019, another two in 2020, and one more in 2021. As of April 2021, the detector consists of eight clusters, occupying a water volume of ~0.4 km³. The present construction plan for the period 2022 - 2024 envisages the deployment of six additional clusters. Figure 19 shows a schematic view of the 2021 configuration and of one cluster.

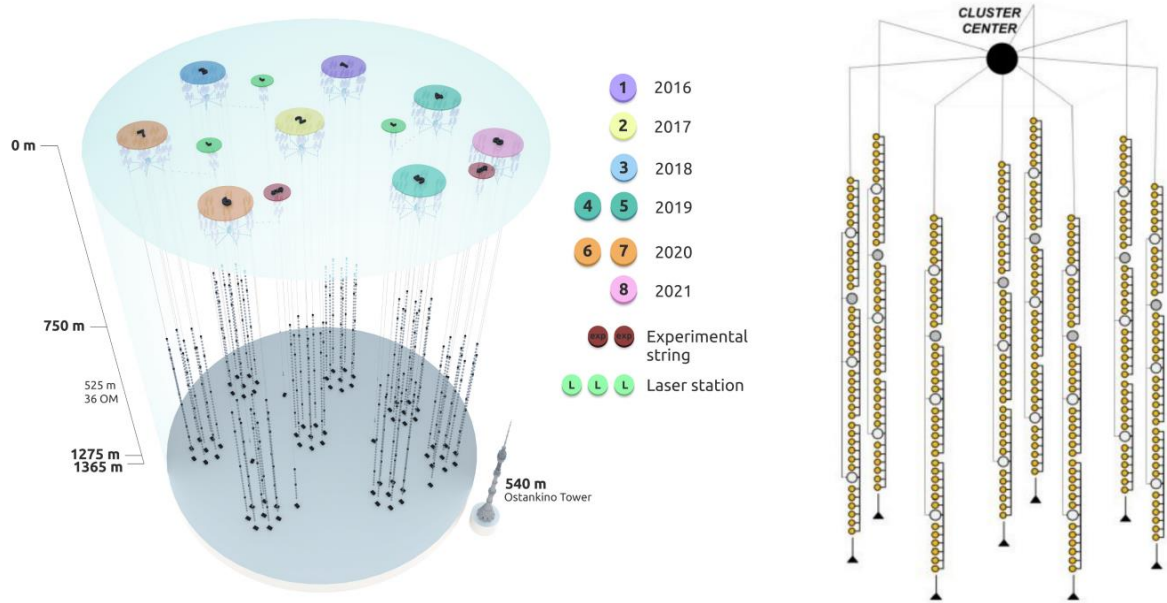

**Figure 19:** *Left: Schematic view of the Baikal-GVD detector [68]. The yearly progression of the detector deployment is shown in the legend. Right: The Baikal-GVD cluster layout (vertically compressed to make details better visible).*

GVD differs from NT200 not only in size. Most importantly, there are worlds between the reliability of the old and the new detector, due to the design of the electronic and mechanical components, to a strict quality management and to extensive testing under laboratory conditions. Presently the failure rate is still higher than for IceCube and even ANTARES, but this is mitigated by the fact that failed components can easily be hauled up and repaired in winter.

While the NT200 project was very strongly dominated by INR, GVD would not have been possible unless JINR Dubna would have taken a comparable role in construction and the dominant role in funding. JINR has taken on the assembly of the optical modules, the data acquisition and data storage and has increased its role in data analysis. Also, the solid extension of the shore station would not have been possible without the support of JINR.

Due to the focus on construction, the project still lags a bit behind of what a collaboration with more manpower could have achieved. However, first results are going to appear. Figure 20, left, shows the zenith angle distribution of 44 almost vertically upward moving tracks from neutrino interactions [68, 69]. Their number as well as the angular distribution are in perfect agreement with Monte Carlo simulations of atmospheric neutrino interactions. Atmospheric neutrinos are the standard candle for understanding the functioning of the detector, therefore these very first results must not be underestimated. Another important result lays along the proven strategy to search for high-energy cascades with energies of several tens of TeV and higher. Figure 20, right, shows the event display of a cascade which has deposited (91±10) TeV. The cascade is upward-going which excludes that it stems from a bremsstrahlung event along a (non-detected) down-going muon. At these energies the chances to be of atmospheric origin are only 1/3, i.e. this event is very likely the interaction of a cosmic neutrino [70].

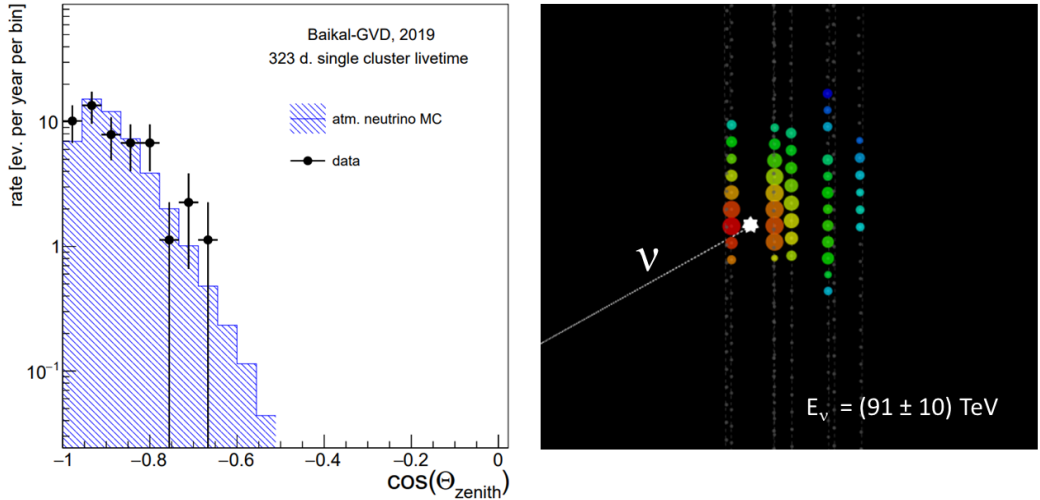

**Figure 20:** *Preliminary GVD results. Left: zenith angle distribution of 44 upward going track-like neutrino events extracted from 323 cluster ×days in 2019. Right: An upward directed cascade with 35 hit PMTs and a reconstructed energy of (91±10) TeV.*

A remarkable event of the last years was the merger of two neutron stars detected in 2017 via its gravitational waves. This event GW170817 happened in the galaxy NGC4993. Two GVD clusters were operational in 2017. Figure 21 [71] shows the neutrino fluence limits derived from GVD data compared to limits from ANTARES and IceCube. Only cascade data have been used for this analysis. Today, with eight clusters, and assuming that not only cascade but also track data would be used, GVD limits would come close to those from IceCube.

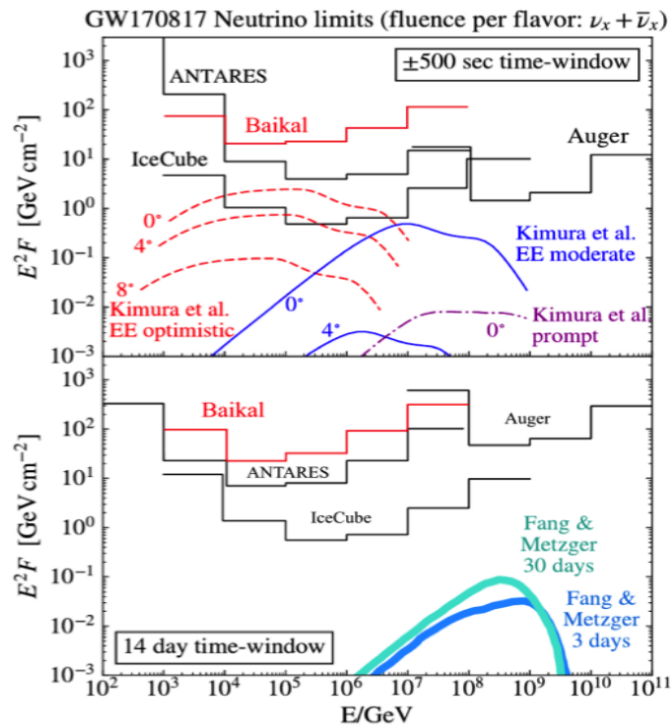

**Figure 21:** *Upper limits at 90% C.L. on the fluence of neutrinos associated with GW170817 for prompt and delayed emission time. Baikal data have been obtained with a configuration only a quarter of what has been installed presently, and using only cascade events. Limits are compared to various predictions (see [71] for details).*

This is a historical review and not an analysis of future prospects. Yet, I cannot help but be reminded a sentence which I wrote in 2000 for a popular-scientific article: *"It is not clear whether the Baikal Telescope will play a main motif or just a side tune at the concert of future big neutrino telescopes".*

Meanwhile, we know the answer: it is going to play a main motif!

## 4.5 Radio and acoustic detection of neutrinos

The detection of neutrinos via the emitted Cherenkov light of secondary particles is arguably the best method in the energy range up to ~10 PeV. However, spectra of cosmic neutrinos from Active Galactic Nuclei and their jets may extend to much higher energies. Moreover, neutrinos are expected from interactions of cosmic protons with the 3K background radiation. These "GZK interactions" [13,14] lead to neutrinos with energies above 100 PeV [72].

For higher energies, alternative methods must be considered. The reason is the limited range of visible light in water and ice. Technologies for higher energies must make use of signals which propagate with km-scale attenuation and allow for the observation of much larger volumes than those typical for optical neutrino telescopes. These methods use either radio or acoustic signals emitted from particle cascades of extremely high energy.

Coherent Cherenkov radiation at radio frequencies is emitted by electromagnetic cascades. The effect was predicted in 1962 by Gurgen Askaryan [73]. Electrons are swept into the developing shower, which acquires an electric net charge from the added shell electrons. This charge propagates like a relativistic pancake of 1 cm thickness and 10 cm diameter. Each particle emits Cherenkov radiation, with the total signal being the convolution of the overlapping Cherenkov cones. For wavelengths larger than the cascade diameter, coherence is observed and the signal rises proportional to $E_\nu^2$.

Acoustic signals are expected from high energy cascades that deposit energy into the medium via ionization losses which are immediately converted into heat. The effect is a fast expansion, generating a bipolar acoustic pulse with a width of a few tens of microseconds in water or ice. This effect was predicted already in 1957, again by Gurgen Askaryan [74]. The group of Igor Zheleznykh has focused to innovative methods of this kind since the 1970s. Here I want to mention the work on radio neutrino detection in Antarctica (RAMAND project), on hydro-acoustical neutrino detection and on neutrino detection with radio telescopes (RAMHAND project).

The group was the first to propose coherent radio waves as an experimental cosmic ray detection technique in 1984. They performed the first experimental studies and laid the initial groundwork for the development of a radio-wave neutrino detectors [75,76]. Their work included measurements at the Vostok station in Antarctica, including the temperature profile of the ice down to 2 km and measurement of the frequency and temperature dependence of the absorption of radio waves in ice. Between 1985 and 1990, a pilot experiment ("RAMAND", for **R**adio wave **A**ntarctic **M**uon **A**nd **N**eutrino **D**etector [77,78] – see Fig. 21) tested many aspects of neutrino radio detection at Vostok, including critical initial measurements of natural and man-made radio impulse backgrounds in Antarctica. Unfortunately, this nascent effort was terminated in 1991, when the Soviet Union collapsed.

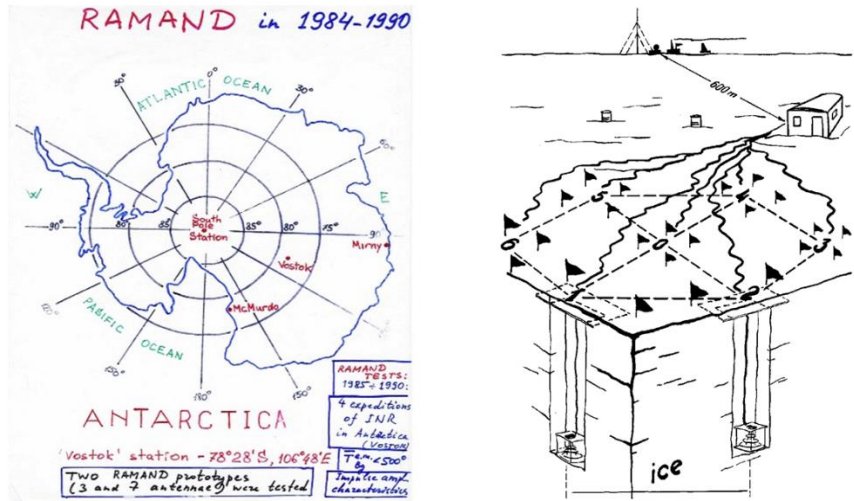

**Figure 21**: *Location and concept of the RAMAND neutrino detector (slides of Igor Zheleznykh). A full RAMAND would have consisted of a large array of radio antennas buried at moderate depths.*

This pioneering concept has been realized at the Amundsen-Scott station in 1995 in the form of a medium-sized radio array named RICE [79]. In recent years two collaborations have deployed prototype detectors for a future array of ~100 km² and deployed them on the Ross ice shelf and at the South Pole. Such a detector is conceived to be part of IceCube-Gen2, a future detector with 8 km³ covered with optical detectors, embedded in a 100 km² radio array.

The hydro-acoustical method has been considered since the early days of the DUMAND project. During the cruise with the Vityaz vessel in 1991 [80], INR researchers performed a series of hydro-acoustic measurements (project SADCO [81]). They also attempted to use former military hydroacoustic arrays for test explorations. The one was a 2400-hydrophone array off the coast of Kamchatka, the second a 132-hydrophone-array in the Black Sea [81]. Due to security and financial reasons and since naturally the focus in the Mediterranean Sea was on the optical detection method, these projects have not been realized. The most advanced project for hydroacoustic detection was the SPATS detector deployed at the South Pole, as a sub-project of DESY/Zeuthen within IceCube. It led to three publications in peer-reviewed journals but was also terminated since the attenuation length turned out to be smaller than expected and the energy threshold too high (see e.g. [82]). Instead, the IceCube-collaboration now focuses to the radio-wave detection at energies > 10 PeV.

The third of the mentioned methods became known under the name RAMHAND (Radio Astronomical Method of Hadron and Neutrino Detection). In the 1980s, the INR group together with researchers from the Lebedev Institute proposed to use the Moon as a giant target and ground-based radio telescopes to detect the long-wavelength tail of Cherenkov bursts from neutrino interactions in the lunar regolith [83]. This method has then been applied by several radio telescopes worldwide, including the Russian Kalyazin telescope [84], leading to flux limits in the $10^{20}$ eV range.

## 5. Review and Outlook

The Institute for Nuclear Research can look back to a great history in high-energy astroparticle physics. Many basic ideas in the field have been formulated by INR researchers: the concept of underwater detection of neutrinos, the concept to use Cherenkov detectors for gamma-ray astronomy, the determination of a high-energy cut-off of cosmic rays and the calculation of the corresponding neutrino flux, the determination of an upper limit for the diffuse neutrino flux, the radio detection of neutrinos in ice and with radio telescopes. With the detectors in the Baksan valley and at Lake Baikal, INR has played also experimentally a pioneering role in the field.

For the future of high-energy cosmic physics, INR is well positioned, in particular with three top projects: Baikal-GVD is one of the worldwide three high-energy neutrino telescopes and at present the largest at the Northern hemisphere. The Telescope Array defines the frontier of cosmic-ray physics at the very highest energies, together with the Pierre Auger Observatory. And, last not least, the LHAASO Observatory in China could become a quantum leap in gamma-ray astronomy. The participation in these three projects alone will allow for a very strong role of INR in multi-messenger observations.

I wish my INR colleagues exciting years to come and success in charting the high-energy cosmos.

## Acknowledgements

I am indebted to Grigori Domogatsky, Alexander Lidvansky, Valery Rubakov, Grigory Rubtsov, Yuri Stenkin, Sergey Troitsky and Igor Zhelesnykh for helpful remarks and corrections of the text. I thank my colleagues of the Baikal collaboration for sharing my first steps in neutrino astronomy and for being friends since 33 years.

## Addendum

This article has covered the history, the achievements and the future plans of INR for *high-energy* astroparticle physics. That excludes neutrinos generated in fusion processes (like neutrinos from the Sun) or by thermal processes (like neutrinos from Supernova bursts). On the other hand, these are frontiers where INR has – again – written history. That's why I would like to add a short addendum on INR's role in solar and Supernova neutrino physics.

February 23, 1987 marks the discovery of the Supernova SN1987 A, the first supernova of the century which was close enough to lead to a measurable neutrino signal on Earth. Alexeyev's BUST group announced five neutrino candidates from SN1987 A [85] in timely coincidence with observations by Kamiokande (Japan) and IMB (USA) [86, 87]. The two dozen neutrinos detected at 7:35 UT by these three detectors became a bonanza for results on particle physics and supernova physics (neutrino mass, neutrino lifetime, neutrino magnetic moment and charge, temperature inside the neutron star etc.).

While the signals in Kamiokande, IMB and BUST coincided in time, about 4½ hours earlier, at 2:52 UT, another excess of neutrinos was recorded in LSD (Liquid Scintillation Detector) [88]. LSD was located under the Mont Blanc and operated by a Russian-Italian collaboration, with a leading role of the INR group chaired by Olga Ryazhskaya. The chance probability to observe the excess of five events within seven seconds was extremely low. Sophisticated "double-bang" models have been developed, also attempting to explain the fact that neither the first three experiments observed significant excesses at 2:52 nor LSD observed an excess at 7:35 [89,90,99], but the very detailed mechanism underlaying SN1987A presumably must still be considered an open question.

The successor of LSD was installed in the Gran Sasso Laboratory in Italy, again with the INR group as dominant partner. The detector, named LVD for Large Volume Detector, is filled with 1000 tons of liquid scintillator (eleven times more than LSD and five times more than BUST). Its primary purpose was the detection of supernova neutrinos. Regrettably no galactic Supernova happened since 1992 when LVD started operation, but LVD could put a stringent upper limit on the rate of core-collapse and failed supernova explosions out to distances of 25 kpc (about one per decade [91]).

The SAGE experiment in the Baksan Neutrino Laboratory [92,93] was led by Vladmir Gavrin. It was built to measure the capture rate of solar neutrinos in the reaction $\nu_e + {}^{71}Ga \rightarrow {}^{71}Ge + e^-$. Together with the GALLEX experiment [94] in the Gran Sasso Laboratory, it was the first experiment with sensitivity to the proton-proton fusion reaction, $p + p \rightarrow d + e^+ + \nu_e$ which generates most of the Sun's energy and is almost independent on details of the solar model. SAGE contains about 57 tons of Gallium, GALLEX 30 tons. SAGE started data taking in 1990, GALLEX in 1991. While SAGE has been running until about 2 years ago, GALLEX (respectively its second stage GNO) was shut down in 2003. The results of both experiments confirmed that the solar neutrino deficit is *not* due to an incorrect solar model but to neutrino oscillation (letting a small chance to neutrino decays which could be rejected by the SNO experiment in 2001).

The understanding of the solar neutrino results at energies higher than those of the *pp* reaction would not have been possible without the MSW effect (where MSW stands for Mikheyev-Smirnov-Wolfenstein). L. Wolfenstein had shown in 1978 that oscillation parameters of neutrinos are changed in matter [95]. In 1985, Stanislav Mikheyev and Alexei Smirnov (both INR) found that a decrease of the density of matter (as along the path of neutrinos generated in the center of the Sun and travelling outwards) can resonantly enhance the neutrino mixing [96,97]. This effect is strongest for ${}^8B$ solar neutrinos and absent for *pp*-neutrinos.

The Ga-Ge technique which has so successfully served to study solar neutrinos is presently used for the experiment BEST which is going to test the possible existence of a fourth (sterile) neutrino [98]. The long-term future project of the Baksan Laboratory, however, is a huge liquid scintillation detector which is mainly devoted to the study of geoneutrinos and solar neutrinos.